# Electron-Phonon Scattering governs both Ultrafast and Precessional Magnetization Dynamics in Co-Fe Alloys

Ramya Mohan[1], Victor H. Ortiz[2], Luat Vuong[2], Sinisa Coh[2], Richard B. Wilson[1,2*]

[1]*Materials Science & Engineering Program, University of California, Riverside*

[2]*Department of Mechanical Engineering, University of California, Riverside*

*\* Corresponding Author:* rwilson@ucr.edu

**Abstract**

Recent investigations have advanced the understanding of how structure-property relationships in ferromagnetic metal alloys affect the magnetization dynamics on nanosecond time-scales. A similar understanding for magnetization dynamics on femto- to pico-second time-scales does not yet exist. To address this, we perform time-resolved magneto optic Kerr effect (TRMOKE) measurements of magnetization dynamics in Co-Fe alloys on femto- to nano-second regimes. We show that Co-Fe compositions that exhibit low Gilbert damping parameters also feature prolonged ultrafast demagnetization upon photoexcitation. We analyze our experimental TR-MOKE data with the three-temperature-model (3TM) and the Landau-Lifshitz-Gilbert equation. These analyses reveal a strong compositional dependence of the dynamics across all time-scales on the strength of electron-phonon interactions. Our findings are beneficial to the spintronics and magnonics community, and will aid in the quest for energy-efficient magnetic storage applications.

**Introduction**

Laser excitation of a magnetic metal causes energy to cascade from photoexcited electrons into spin and vibrational degrees of freedom[1–3]. In ferromagnetic 3d transition metals such as Fe, Co, and Ni, the rapid increase in thermal energy stored by spin degrees of freedom causes femtosecond quenching of the magnetization[2,3], followed by a partial recover over the next few picoseconds. Subsequently, on nanosecond time-scales, a temperature induced change in equilibrium properties causes oscillatory precessions of the magnetic moment.

Both ultrafast and precessional magnetization dynamics involve energy exchange between magnetic and vibrational degrees of freedom. The energy exchange is mediated by quasi-particle interactions. The strength of quasi-particle interactions in a ferromagnet depends on electronic band structure[4,5]. In 3d ferromagnetic alloys, the electronic energy bands near the Fermi-level vary strongly with composition[6]. Several recent investigations of nanosecond precessional dynamics in ferromagnetic alloys have explored the relationship between electronic band structure, quasi-particle interactions, and magnetic damping[6–8]. Schoen *et al.* report an intrinsic damping parameter less than $10^{-3}$ for $Co_{0.25}Fe_{0.75}$[6], which is unusually low for a metal. They conclude that the low damping in $Co_{0.25}Fe_{0.75}$ is a result of a minimization in the density of states at the Fermi-level, which decreases the rate of electron-phonon scattering.

Researchers have not yet reached a unified understanding of how quasi-particle interactions govern the magnetization dynamics in the femtosecond regime[2,9–15]. Some studies have hypothesized that spin-flips caused by electron-phonon interactions are key drivers of femtosecond magnetization dynamics[9,11]. Other experimental and theoretical studies have explored the importance of electron-magnon interactions[12–15]. Encouraged by the recent advances in the materials science of nanosecond precessional dynamics[6–8], we study the compositional dependence of ultrafast magnetization dynamics in Co-Fe alloys. Our study's goal is to understand the relationship between electronic band structure, quasi-particle interactions, and femto-magnetism properties of ferromagnetic metal alloys.

We perform time-resolved magneto optic Kerr effect (TR-MOKE) measurements to characterize the magnetization dynamics of thin $Co_xFe_{1-x}$ alloy films (capped and seeded with Ta/Cu layers on a sapphire substrate) on femto- to nanosecond time-scales. See Methods for details on sample geometry. We observe that the ultrafast magnetization dynamics are a strong function of Co-concentration, see Figure. 1a. The ultrafast dynamics of $Co_xFe_{1-x}$ differ most significantly from those of Co and Fe at a composition of $x = 0.25$. We also analyze the time-resolved macroscopic precessional dynamics and report the effective damping parameter of our samples, see Figure 2a. After linewidth analyses, for $Co_xFe_{1-x}$, we observe that the Gilbert damping parameter varies from $3.6 \times 10^{-3}$ to $5.6 \times 10^{-3}$ for compositions between $x = 0$ and 1, with a minimum value of $1.5 \times 10^{-3}$ at $x = 0.25$, in good agreement with previously reported results, see Figure 3b.

To determine the strength and composition dependence of electron-magnon and electron-phonon quasi-particle interactions, we analyze our ultrafast magnetization dynamics data with a three-temperature-model (3TM)[2,16]. Our results reveal a strong compositional dependence of the electron-phonon energy transfer coefficient, $g_{ep}$, suggesting that the variation in the ultrafast dynamics in $Co_xFe_{1-x}$ alloys occurs primarily due to electron-phonon scattering. We draw this conclusion because the value of $g_{ep}$ depends on the rate of phonon emission by hot electrons [17]. Electron-phonon scattering is also predicted to govern the damping of nanosecond precessional dynamics [6,18,19]. Therefore, our results demonstrate that the same microscopic electron-phonon interactions responsible for Gilbert damping also play a dominant role in femto-magnetism properties of ferromagnetic alloys.

**Results**

**Ultrafast Magnetization Dynamics**

We plot the normalized ultrafast magnetization dynamics response, $\Delta M(t)$, for Co, Fe, and $Co_{0.25}Fe_{0.75}$ as a function of time delay in Figure. 1a. Data for the rest of the Co-Fe compositions are plotted in Supplementary Figure 1. All our measurements were performed with an incident laser fluence less than ~15 J/m$^2$. This is a sufficiently small fluence for the dynamics in our experiments to follow a linear regime. In other words, decreasing the incident fluence by a factor of two decreases the optical signal by a factor of two, but does not change the time-dependence of the signal.

We use a polar TR-MOKE configuration to measure the ultrafast magnetization dynamics at femtosecond time delays. A schematic of our experimental setup is shown in Supplementary Figure 2a. We apply an external 2.2 Tesla (T) field perpendicular to the plane of the sample using an electromagnet (GMW 3480). This external field is strong enough to effectively overcome the in-plane shape anisotropy of the Co-Fe alloys and saturate the moment in the out-of-plane direction. Since the equilibrium orientation of the moment is in the out-of-plane direction, both, before and after laser irradiation, this geometry allows us to quantify the femtosecond demagnetization response of the Co-Fe alloys, without the presence of macroscopic precessional dynamics, see schematic in Figure 1b.

Upon excitation with the pump pulse, the magnetic moment decreases on a sub-picosecond time-scale due to the flow of energy from electrons to magnons[2,3,16,20,21]. Then, on picosecond time-scales, the magnetization partially recovers as energy is transferred to the lattice and temperature gradients across the film thickness relax. After a few picoseconds, the magnetic film reaches a new equilibrium at an elevated temperature. Ultrafast dynamics with sub-picosecond demagnetization followed by picosecond re-magnetization are commonly categorized as "type I" dynamics, and are characteristic of 3d ferromagnetic metals such as Fe, Co, and Ni[9].

To elucidate how the de- and re-magnetization dynamics change with composition, we define two data descriptors: $\tau_D$ and $R$. We define the demagnetization time, $\tau_D$, as the delay time where d$\Delta M(t)$/d$t$ reaches its maximum value. We define $R$ as the ratio of the maximum of $\Delta M(t)$ to $\Delta M(t \approx 10\text{ps})$. We plot $\tau_D$ and $R$ as a function of composition in Figure 3a. $\tau_D$ varies weakly with composition and has a minimum value of 40 fs at $x = 0.25$. In contrast, we observe that $R$ varies strongly with composition and is a maximum of 4 at $x = 0.25$.

**Nanosecond Precessional Dynamics**

We show measurements of the macroscopic precessional dynamics of Fe, Co, and Co$_{0.25}$Fe$_{0.75}$ in Figure 2a. Data for the other Co-Fe compositions are plotted in Supplementary Figure 3. We use a polar TR-MOKE experimental setup, with an obliquely angled external magnetic field, to measure the macroscopic precessional dynamics of our samples. A schematic of our experimental setup is shown in Supplementary Figure 2b. Tilting the electromagnet to an angle of 11°, with respect to the plane of the sample, allows us to apply a canted external magnetic field so that the magnetic moment has an out-of-plane component. The equilibrium orientation of the moment depends on the balancing between the applied external field and the thin-film shape anisotropy field. The shape anisotropy field in the z-direction is proportional to the out-of-plane component of the magnetic moment. Upon heating, the total magnetic moment decreases. This decrease results in an ultrafast change to the out-of-plane anisotropy field and equilibrium orientation. As a result, the magnetic moment will precess to a new equilibrium orientation, see schematic in Figure 2b. Our polar TR-MOKE setup detects changes in the out-of-plane moment, so we can sensitively measure the frequency and amplitude of the precessional dynamics.

We collect between 6 and 12 TR-MOKE scans of precessional dynamics for each sample. Each of these scans is collected with a different applied external magnetic field, ranging from 0.2 T to 2.2 T. The TR-MOKE signals include precessional dynamics in addition with a background related to temperature-induced demagnetization. To analyze the precessional dynamics, we subtract the background with a biexponential decay function. We fit the resulting dataset with a damped harmonic function, $V(t) = A \sin(\omega t + \emptyset) \exp(-t/\tau)$. Our fits yield unique values of A (amplitude), $\emptyset$ (the initial phase of the oscillation), T (period), and $\tau$ (the exponential decay time of the precession). Using these values, we determine the effective dimensionless damping parameter, $\alpha_{eff} = \omega \cdot \tau^{-1}$.

The resonance frequency is a function of applied external magnetic field and magnetic moment, $\omega = \gamma \sqrt{H_{eff}(H_{eff} + \mu_0 M_s)}$. Here, $\gamma$ is the gyromagnetic ratio, $\mu_0$ is the vacuum permeability, $H_{eff}$ is the out-of-plane component of the external magnetic field as measured by a Hall probe, and $M_s$ is the saturation magnetization of the sample. We derive the magnetic moment of the sample by treating $M_s$ as a fit parameter. We also perform VSM measurements of the moment of some of the samples and find that the magnetic moment obtained is in good agreement with the value that we derive by fitting our precessional dynamics data. See Supplementary Figure 4 for more details.

The effective damping parameter $\alpha_{eff}$ that we deduce from our precessional dynamics measurements includes effects from damping and inhomogeneous broadening. The effect of inhomogeneous broadening is independent of the applied field at high frequencies[22]. To obtain the Gilbert damping parameter intrinsic to the sample geometry (not intrinsic to the material), we plot the effective linewidth, $\alpha_{eff} \cdot f$, as a function of frequency, and linearly fit to the equation, $\alpha_{eff} \cdot f = \alpha \cdot f + \Delta H$, where $\Delta H$ is the inhomogeneous broadening component and $\alpha$ is the Gilbert damping parameter. Further details can be found in Supplementary Figure 5.

In contrast to prior investigations that performed FMR measurements in the frequency range from 16-18 GHz[8] and 40 GHz[6], our TR-MOKE experimental setup allows us to study dynamics at frequencies as large as 90 GHz. At such high frequency, we can be confident that our measured Gilbert damping parameter is dominated by the intrinsic linewidth over inhomogeneous broadening effects.

The Gilbert damping parameter we observe of $\alpha = 1.5 \times 10^{-3}$ for $Co_{0.25}Fe_{0.75}$ is amongst the lowest ever reported for a ferromagnetic metal. Schoen et al. report $\alpha = 2.1 \times 10^{-3}$ for $Co_{0.25}Fe_{0.75}$. After accounting for radiative and spin-pumping contributions, they estimate an intrinsic damping parameter for $Co_{0.25}Fe_{0.75}$ to be $\alpha_{int} = 5 \times 10^{-4}$. Lee et al.[8] performed FMR measurements of $Co_{0.25}Fe_{0.75}$ epitaxial films and report $\alpha = 1.4 \times 10^{-3}$. Wei et al. report $\alpha =$ $1.5 \times 10^{-3}$ for Fe$_{0.75}$Al$_{0.25}$ films [7]. We note that our measured damping parameter likely includes significant contributions from spin-pumping into the adjoining Ta/Cu layers, but we did not experimentally examine the effects of spin-pumping in our samples.

**Analysis and Discussion**

The comparison of $R$ and $\alpha$ in Figure 3a and Figure 3b reveals that the two quantities depend on composition in a similar manner. $R$ is at a maximum and $\alpha$ is at a minimum at $x = 0.25$. Fe and Co$_x$Fe$_{1-x}$ alloys with $x \geq 0.5$ have small $R$ and high $\alpha$. Alternatively, Co$_x$Fe$_{1-x}$ alloys with $0.1 < x < 0.5$ have both high $R$ and low $\alpha$. To confirm this correlation, we performed a hierarchical cluster analysis of the raw data at both femtosecond and nanosecond time-scales. The clustering algorithm divides the Co-Fe alloys into groups based on similarities in the dynamics data. The clustering results as a function of composition are nearly identical when based on the femto-/pico-second time-scale data vs. the nanosecond time-scale data. We include further details on the clustering analysis in Supplementary Note 1 and Supplementary Figure 6.

We now explain the correlation between ultrafast and precessional dynamics by considering how electronic scattering processes depend on composition. Similar to prior studies of damping in Co-Fe alloys[6,7,23], our results for $\alpha$ vs. $x$ are in good agreement with the "breathing Fermi surface" model for damping[24]. In this model, spin-orbit coupling causes the Fermi-level to shift with the precessions of the magnetic moment[25]. A shift in the equilibrium Fermi-level leads to a nonequilibrium electron population. As the Fermi-level repopulates, intra-band electron-phonon scattering transfers energy to the lattice. The "breathing Fermi surface" model predicts that the damping parameter is directly proportional to $D(\varepsilon_f)$, because more electronic states near $\varepsilon_f$ leads to higher rates of electron-phonon scattering. We observe that the $\alpha$ value for Co$_{0.25}$Fe$_{0.75}$ is ~2.5x lower than $\alpha$ for Fe. Density functional theory predicts a ~2x difference in $D(\varepsilon_f)$ for Co$_{0.25}$Fe$_{0.75}$ vs. Fe, see Supplementary Note 2 or Ref.[6]. Therefore, like prior studies of Co-Fe alloys[6,7,23], we conclude that intra-band electron-phonon scattering governs precessional damping.

To better understand how composition affects electron-magnon and electron-phonon energy transfer mechanisms, we analyze our $\Delta M(t)$ data with a phenomenological three temperature model (3TM), see Figure 4. The 3TM describes how heat flows between electrons, phonons, and magnons after laser excitation of the Co-Fe sample. (See Methods for additional details.) The 3TM predicts that $\tau_D$ depends on two groupings of model parameters: $\tau_{em} \approx C_m/g_{em}$ and $\tau_{ep} \approx C_e/g_{ep}$. Here $C_m$ and $C_e$ are the magnon and electron heat-capacity per unit volume, and $g_{em}$ and $g_{ep}$ are the energy transfer coefficients from electrons to magnons and phonons, respectively. We estimate values for $C_e$ vs. composition using the Sommerfeld model together with the electronic density of states vs. composition reported in Ref.[6]. The 3TM also predicts that the parameter $R$ is determined by the following grouping of parameters: $R = C_p g_{em}/C_m g_{ep}$ [16], where $C_p$ is the phonon heat-capacity per unit volume. We assume that the value of $C_p$ is 3.75

MJ m$^{-3}$ K$^{-1}$ for Co, Fe and Co-Fe alloys. With these estimates for $C_e$ and $C_p$, and other relevant model parameters, summarized in Supplementary Table 1, we can deduce unique values for $C_m/g_{em}$ and $C_p/g_{ep}$ as a function of composition from our TR-MOKE data, see Figure 4b.

Based on our 3TM analysis, we conclude that the strong composition dependence of $R$ is due to the composition dependence of $g_{ep}$. Boltzmann rate-equation modelling of the nonequilibrium electron dynamics after photoexcitation predicts that the electron-phonon energy-transfer coefficient is $g_{ep} = [\pi\hbar k_B D(\varepsilon_F)]\lambda\langle\omega^2\rangle$ [5]. Here, $\lambda\langle\omega^2\rangle$ is the second frequency moment of the Eliashberg function and is a measure of the strength of electron-phonon interactions. Most of the compositional dependence we observe in $g_{ep}$ is explained by the compositional dependence of $D(\varepsilon_f)$. To show this, we include a prediction for $g_{ep}$ in Figure 4b. Our prediction uses the $D(\varepsilon_f)$ vs. $x$ reported in[6] and treats $\lambda\langle\omega^2\rangle$ as a composition independent fit parameter. We find $\lambda\langle\omega^2\rangle = 260$ meV$^2$ provides an excellent fit to our data. The best-fit value for $\lambda\langle\omega^2\rangle$ is in good agreement with $\lambda\langle\omega^2\rangle \approx \lambda_R \Theta_D^2/2 = 280$ meV$^2$. Here, $\lambda_R$ is derived from electrical resistivity data for Fe [26], and $\Theta_D = 470 K$ is the Debye temperature of Fe.

Before beginning our experimental study, we hypothesized that the energy transfer coefficient between electrons and magnons, $g_{em}$, would be correlated with the phase-space for electron-magnon scattering. We expected the phase-space for electron-magnon scattering to be a strong function of band-structure near the Fermi-level [12–15]. We also expected the phase-space to be minimized at a composition of $x = 0.25$, because of the minimum in the density of states at the fermi-level. To explore how the phase-space for electron-magnon scattering depends on composition, we performed density functional theory calculations for the electronic band structure with $x = 0$ and $x = 0.25$, see Supplementary Note 2. Our DFT calculations suggest that the phase-space for electron-magnon scattering is an order of magnitude higher for $x = 0$ vs. 0.25. However, we do not see evidence that this large theoretical difference in electron-magnon scattering phase-space affects ultrafast dynamics. The time-scale for magnons to heat up after photoexcitation, $\tau_{em} \approx C_m/g_{em}$, decreases monotonically with increasing $x$, and does display structure near $x \sim 0.25$.

Several theoretical models predict a strong correlation between $\tau_D$ and $\alpha_{int}$. For example, Koopmans *et al.* predicts $\tau_D$ will be inversely proportional to $\alpha$ by assuming that the dissipative processes responsible for damping also drive ultrafast demagnetization [27]. Alternatively, Fähnle *et al.* predicts that $\tau_D$ should be proportional to $\alpha_{int}$ [28]. In our experiments on Co-Fe thin films, we observe only a weak correlation between $\tau_D$ and $\alpha_{int}$. While $\alpha_{int}$ varies with composition by a factor of three, $\tau_D$ for 8 of the 9 compositions we study fall within 20% of 75 fs. The $\tau_D$ value we obtained for Fe (= 76 fs) agrees well with experimental results reported in [9,12,29].

**Conclusions**

We have measured the magnetization dynamics of $Co_xFe_{1-x}$ thin-films, and we observe that both ultrafast and precessional dynamics of $Co_{0.25}Fe_{0.75}$ differ significantly from Co and Fe. When the moment of $Co_{0.25}Fe_{0.75}$ is driven away from its equilibrium orientation, the time-scale for the moment to return to equilibrium is 3-4x as long as for Fe or Co. Similarly, when spins of $Co_{0.25}Fe_{0.75}$ are driven into a nonequilibrium state by ultrafast laser heating, the time-scale for thermalization with the lattice is 2-3x as long as for Fe or Co. Through 3TM analyses, we demonstrate that this occurs primarily due to the effect of the electronic band-structure on electron-phonon interactions, consistent with the "breathing Fermi surface" theory. Our findings are of fundamental importance to the field of ultrafast magnetism, which seeks to control magnetic order on femto- to picosecond time-scales. Such control requires a thorough understanding of how and why energy is exchanged between electronic, spin, and vibrational degrees of freedom. Prior studies have shown that $g_{ep}$ is correlated with a wide range of physical properties, e.g the superconducting transition temperature[30], electrical resistivity [26], photoelectron emission[31], and the laser fluence required for ablation[32]. To our knowledge, our study provides the first demonstration that $g_{ep}$ in ferromagnetic metals is also correlated to the Gilbert damping parameter $\alpha$.

Our findings also have implications for the ongoing search for magnetic materials with ultrafast magnetic switching functionality. Atomistic spin dynamics simulations predict that the energy required for ultrafast electrical or optical switching of rare-earth ferromagnetic alloys, e.g. GdFeCo, is governed by the electron-phonon energy transfer coefficient[33]. To date, most studies aimed at exploring the materials science of ultrafast switching have used alloy composition as a way to control magnetic properties [34–37]. Our work suggests an alternative strategy for reducing the energy requirements for ultrafast magnetic switching. The alloy composition should be chosen to minimize the electronic density of states at the Fermi-level. Such metals will have lower electron-phonon energy transfer coefficients, and therefore more energy efficient ultrafast switching [33].

Finally, our findings offer a new route for discovering ferromagnetic materials with ultra-low damping as a result of low $g_{ep}$. Current methods for identifying low damping materials involve labor-intensive ferromagnetic resonance measurements of one alloy composition at a time. Alternatively, high-throughput localized measurements of ultrafast demagnetization dynamics of samples produced using combinatorial techniques[38] would allow promising alloy compounds with weak electron-phonon interactions to be rapidly identified [39–41].

## Materials and Methods

### Sample Preparation

We sputter deposit the Co-Fe samples onto sapphire substrates with a direct current (DC) magnetron sputtering system (Orion, AJA International). The base pressure prior to deposition is less than $3.5 \times 10^{-7}$ torr. We sputter with an Argon pressure of $\sim 3.5 \times 10^{-3}$ torr. The geometry of the samples is sapphire/Ta(2nm)/Cu(3nm)/Co$_x$Fe$_{1-x}$(15nm)/Cu(3nm)/Ta(1nm). The Co$_x$Fe$_{1-x}$ layer is deposited by co-sputtering two 4N purity Co and Fe targets at different powers. We chose this film geometry to mimic the samples in Ref.[6] which demonstrated low damping at $x = 0.25$.

To ensure an accurate thickness of each layer in our samples, we calibrate the deposition rates of each metal by sputtering individual Co, Fe, Ta, and Cu films onto SiO$_2$/Si substrates and/or BK-7 glass substrates. We use picosecond acoustics[42] and time-domain thermo-reflectance (TDTR) measurements[43,44] to determine the thicknesses of these individual films. We validate the composition of the Co-Fe alloy layer by performing Energy Dispersive X-Ray Spectroscopy (EDS) analyses with a scanning electron microscope (FEI Nova NanoSEM 450) at an operating voltage of 15 kV and working distance of 14 mm. We analyze the EDS data using Aztec Synergy software (Oxford Instruments).

### Time-Resolved MOKE Experimental Setup

We use a pump/probe laser system to perform TR-MOKE measurements of the magnetization dynamics. The pulsed laser is a Ti:sapphire oscillator with an 80 MHz repetition rate. The laser beam is split into a pump and probe beam, that are modulated to frequencies of 10.7 MHz and 200 Hz, respectively. A time-delayed pump beam irradiates the sample surface and heats the metal film. The ultrafast heating causes a change in the magnetic moment. We measure the time-evolution of the magnetic moment by monitoring the polarization of the probe beam reflected off the sample surface. The reflected probe beam's polarization state is affected by the out-of-plane magnetic moment of the sample due to the polar Kerr effect. Additional details about the MOKE experiment set-up are in Ref.[45].

The time-resolution of our experiment is controlled by the convolution of the intensity vs. time of the pump and probe pulses. The wavelength of our pump and probe beams is tunable. Employing a red (900 nm) pump and blue (450 nm) probe yields higher time-resolution capabilities, allowing us to accurately measure the ultrafast magnetization at femtosecond time delays. We measure the full-width-at-half-maximum (FWHM) of the convolution of the pump and probe pulses by performing an inverse Faraday effect (IFE) measurement on Pt. We obtain a FWHM value of 390 fs for the convoluted pulses, and a pulse duration of 210 fs for the 900 nm pump/450 nm probe beam setup. For further details on our IFE measurements and pulse duration calculations, please refer to Supplementary Figure 8.

To investigate the precessional dynamics on longer time-scales, we use a pump and probe wavelength of 783 nm. The pulse duration for this setup is 610 fs due to pulse broadening from a two-tint setup we use to prevent pump light from reaching the balanced detector[45,46].

**Three Temperature Modeling**

To determine the electron, phonon, and magnon energy transfer coefficients, we use the phenomenological three-temperature model (3TM), given by the following set of equations:

$$C_e \frac{dT_e}{dt} = g_{ep}(T_p - T_e) + g_{em}(T_m - T_e) + \Lambda_e \frac{d^2 T_e}{dz^2} + S(z,t) \quad (1)$$

$$C_p \frac{dT_p}{dt} = g_{ep}(T_p - T_e) + \Lambda_p \frac{d^2 T_p}{dz^2} \quad (2)$$

$$C_m \frac{dT_m}{dt} = g_{em}(T_m - T_e) + \Lambda_m \frac{d^2 T_m}{dz^2} \quad (3)$$

$$S(z,t) = S_0 P(t) A(z) \quad (4)$$

Equations 1 – 3 describe the temperature evolution of electrons (e), phonons (p) and magnons (m), as a function of time delay (t). C, T, and $\Lambda$ are the heat capacity per unit volume, temperature, and thermal conductivity, respectively. We use the density of states (DOS) at the Fermi level as a function of Co-concentration[6] to calculate the electronic heat capacity ($C_e$) using the Sommerfeld model. We assume that the phonon-magnon energy transfer is negligible compared to electron-magnon coupling, and thus, neglect $g_{pm}$.

We calculate the laser energy absorption by electrons (S), as a function of depth (z) and time delay (t), as described in Equation 4. The terms P(t) and A(z) denote the time-dependent laser pulse intensity and the optical absorption profile as a function of stack thickness. We calculate A(z) using the refractive indices of each metal constituent of the stack[47–49]. The material parameters that are used to numerically solve equations 1 – 4 are listed in Supplementary Table 1.

Figures:

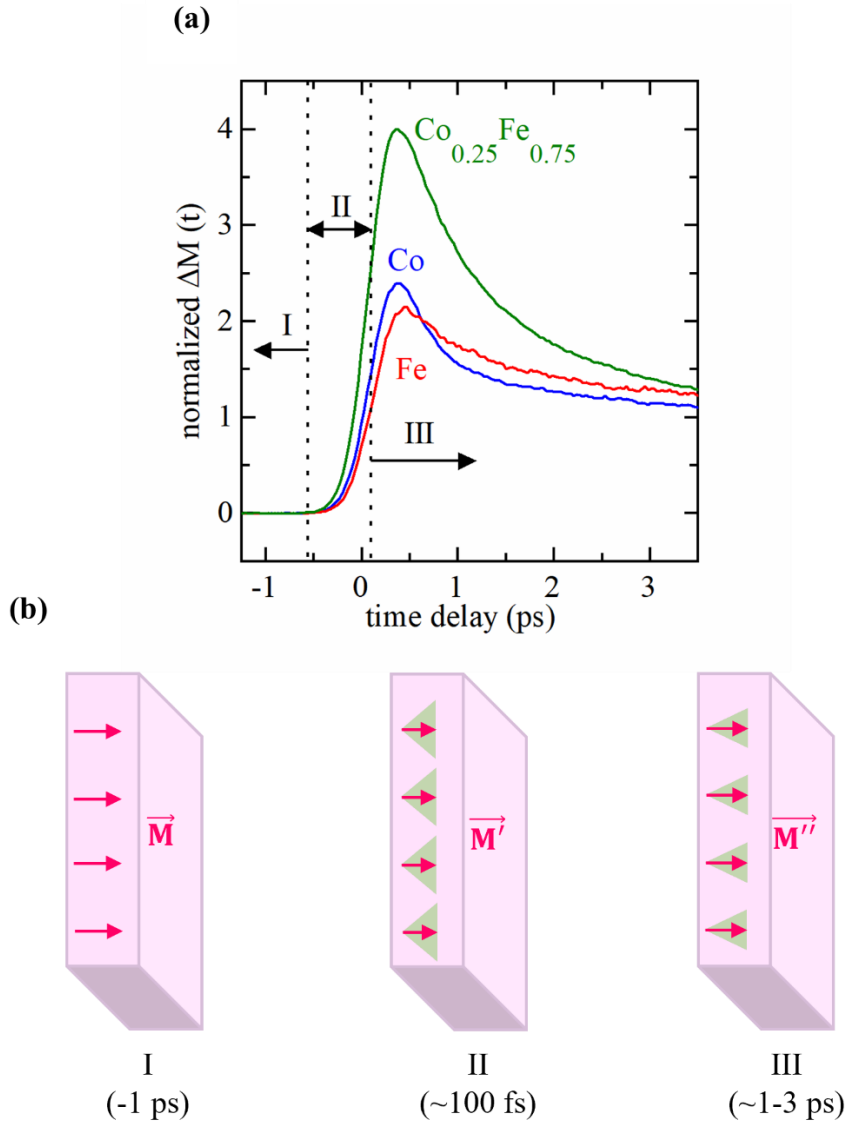

**Figure 1. Ultrafast magnetization dynamics of Co, Fe, and Co$_{0.25}$Fe$_{0.75}$ thin films (a)** Polar TR-MOKE data showing ultrafast demagnetization behavior at short delay times. **(b)** Schematic illustration of the three phases of an ultrafast magnetization dynamics experiment. Stage I: A large external magnetic field oriented normal to the plane of the sample leads to an equilibrium moment, $\vec{M}$ in the out-of-plane direction. Stage II: Upon heating with a pump beam, ultrafast demagnetization ($\vec{M'}$) occurs within ~100s of fs. Energy from hot electrons is transferred to the magnons, increasing the amplitude of precession. Stage III: Over the next few picoseconds, energy is transferred from magnons and electrons to the lattice. Additionally, spatial temperature gradients relax. As a result, magnons cool, i.e. the average precessional amplitude of individual spins decreases. As a result, the magnetization partially recovers to $\vec{M''}$. The time-scale for the partial recovery in stage III depends strongly on the composition.

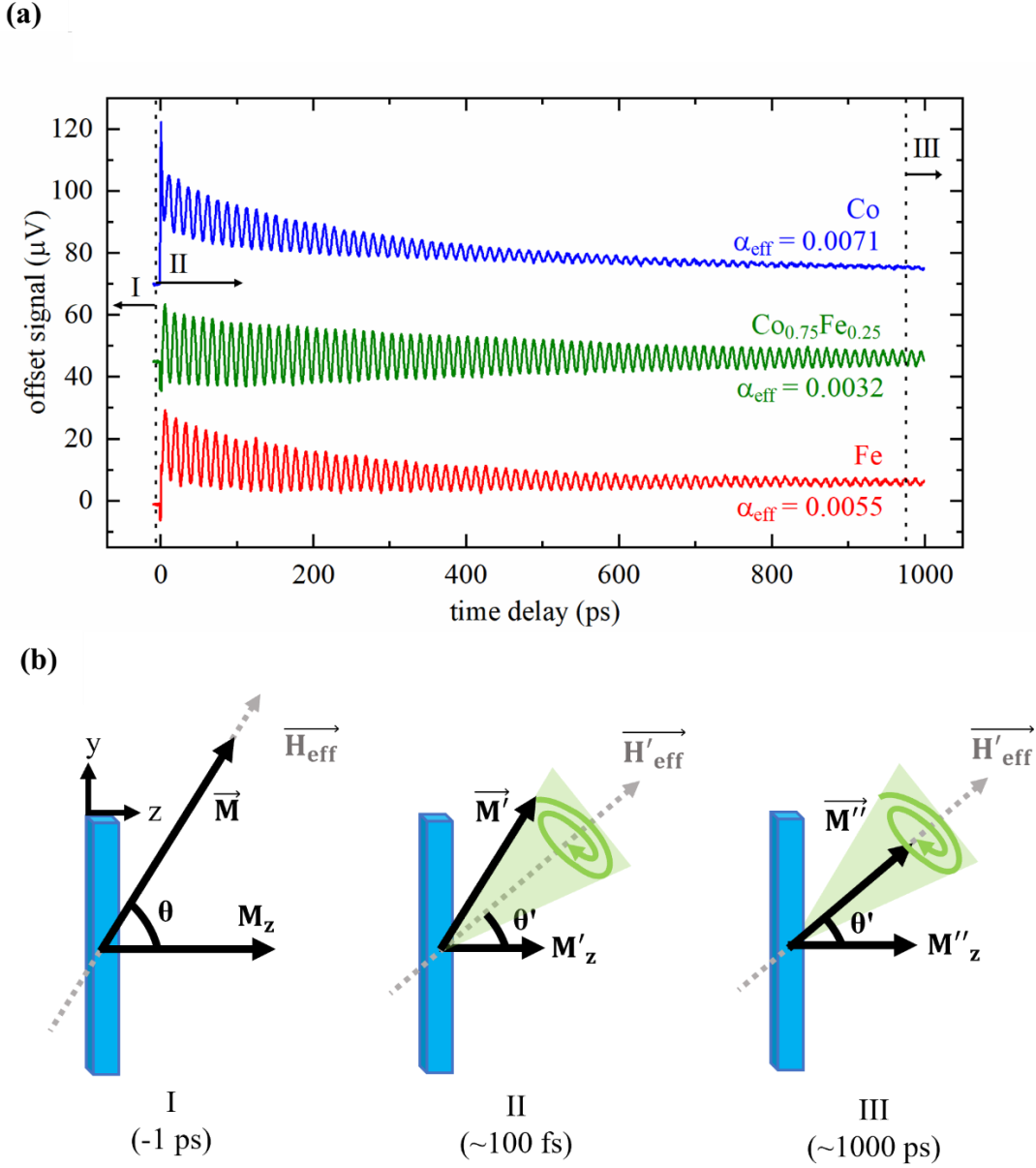

**Figure 2. Precessional dynamics in Co, Fe, and Co$_{0.25}$Fe$_{0.75}$ thin films** **(a)** Polar TR-MOKE data on sub-nanosecond time-scales. **(b)** Illustration of the three stages for precessional dynamics after laser excitation. Stage I: Prior to laser excitation, the presence of a canted external magnetic field, $\overrightarrow{H_{eff}}$, oriented at an angle θ. This results in the orientation of the out-of-plane moments, $\overrightarrow{M_z}$. Stage II: Laser-induced photoexcitation leads to the disorder of the magnetic moment, causing a decay in the net magnetization, denoted by $\overrightarrow{M'}$. The net torque imbalance causes macroscopic precessions of the magnons, towards equilibrium, $\overrightarrow{H'_{eff}}$, over several ~100s of picoseconds. Stage III: Eventually, after ~1 ns, the magnetic moment re-equilibrates to $\overrightarrow{H'_{eff}}$. The lifetime of the magnetic precessions depends on the effective damping parameter, α$_{eff}$. The time-scale for the precessional dynamics to cease (in stage III) depends strongly on composition, and is a maximum for *x = 0.25.*

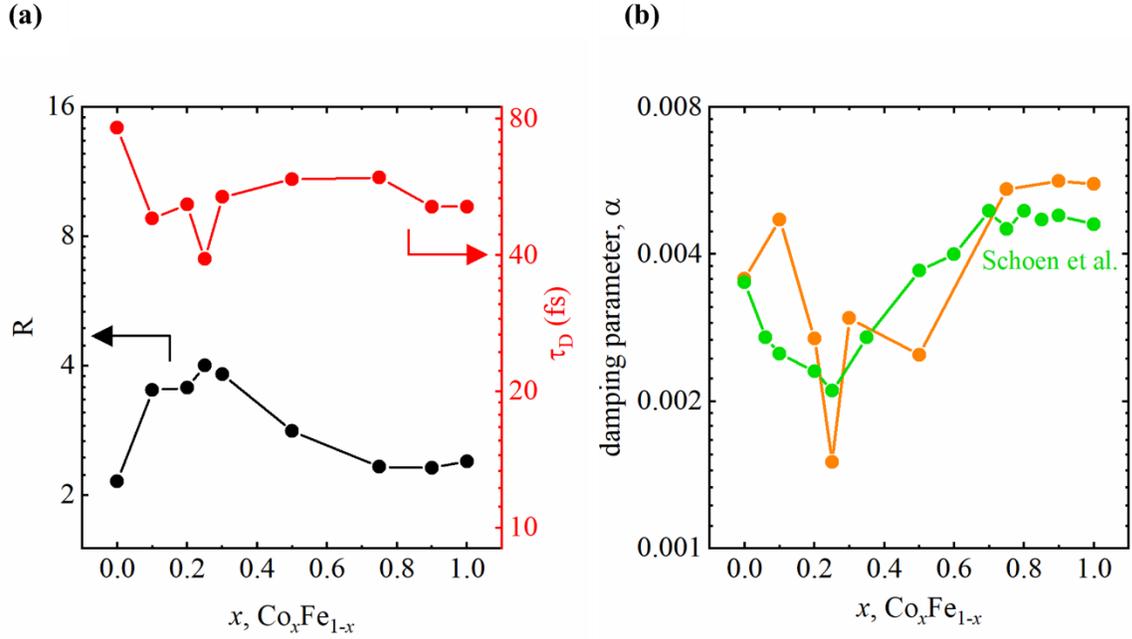

**Figure 3. Compositional dependence of descriptors for the ultrafast dynamics data.** **(a)** $R$ describes the maximum change in the magnetic moment, i.e. how far from equilibrium spin-degrees of freedom are driven after ultrafast excitation. $\tau_D$ describes the lag between zero delay time and demagnetization, as a function of Co-concentration. **(b)** α denotes the Gilbert damping parameter, as a function of Co concentration. Data obtained from our TR-MOKE experiments described in this study (plotted in orange), agree reasonably with data from Ref. [6] (plotted in green). $Co_{0.25}Fe_{0.75}$ features the largest deviation in $R$ and α, when compared to its constituent elements Co and Fe.

(a) 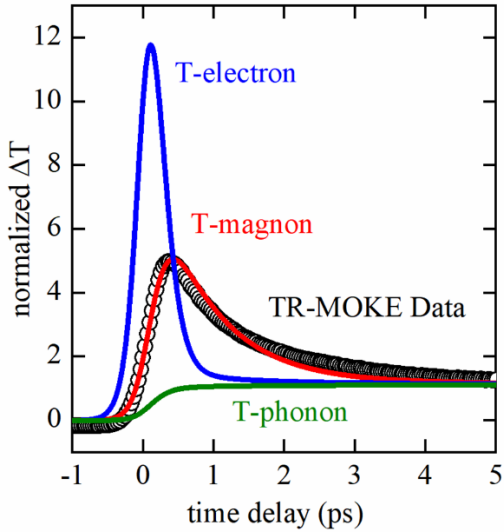

(b) 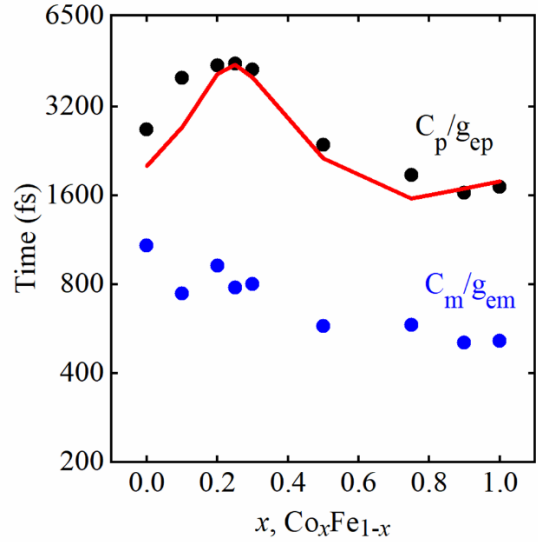

**Figure 4. Analyses of Ultrafast Demagnetization Results using the Three Temperature Model (3TM) in Co-Fe alloys**. (**a**) Polar TR-MOKE dataset of the $Co_{0.25}Fe_{0.75}$ composition (black circles) with best-fit results of the 3TM. The 3TM describes the temperature excursions of the electrons (blue curve), magnons (red curve) and phonons (green curve) after laser excitation. (**b**) We treat $g_{ep}$ and $g_{em}$ as fit parameters when solving the 3TM. Using literature values of $C_p$ and $C_m$ (further details available in Supplementary Table 1), we calculate and plot the electron-phonon ($\tau_{ep}$) and electron-magnon ($\tau_{em}$) relaxation times, as a function of Co-concentration. The red-line is a best-fit value for the electron-phonon relaxation time as a function of composition, with the assumption of a composition-independent value for the electron-phonon coupling parameter $\lambda$.

**References:**

1. Kirilyuk, A., Kimel, A. V & Rasing, T. Ultrafast optical manipulation of magnetic order. *Rev. Mod. Phys.* **82**, 2731 (2010).

2. Beaurepaire, E., Merle, J. C., Daunois, A. & Bigot, J. Y. Ultrafast spin dynamics in ferromagnetic nickel. *Phys. Rev. Lett.* **76**, 4250–4253 (1996).

3. Hellman, F. *et al.* Interface-Induced Phenomena in Magnetism. *Rev. Mod. Phys.* **89**, 025006 (2017).

4. McMillan, W. L. Transition Temperature of Strong-Coupled Superconductors. *Phys. Rev.* **167**, 331–344 (1968).

5. Allen, P. B. Theory of thermal relaxation of electrons in metals. *Phys. Rev. Lett.* **59**, 1460–1463 (1987).

6. Schoen, M. A. W. *et al.* Ultra-low magnetic damping of a metallic ferromagnet. *Nat. Phys.* **12**, 839–842 (2016).

7. Wei, Y. *et al.* Ultralow magnetic damping of a common metallic ferromagnetic film. *Sci. Adv.* **7**, 1–7 (2021).

8. Lee, A. J. *et al.* Metallic ferromagnetic films with magnetic damping under $1.4 \times 10^{-3}$. *Nat. Commun.* **8**, 1–6 (2017).

9. Koopmans, B. *et al.* Explaining the paradoxical diversity of ultrafast laser-induced demagnetization. *Nat. Mater.* **9**, 259–265 (2010).

10. Chen, Z. & Wang, L. W. Role of initial magnetic disorder: A time-dependent ab initio study of ultrafast demagnetization mechanisms. *Sci. Adv.* **5**, eaau8000 (2019).

11. Carva, K., Battiato, M. & Oppeneer, P. M. Ab initio investigation of the Elliott-Yafet electron-phonon mechanism in laser-induced ultrafast demagnetization. *Phys. Rev. Lett.* **107**, 207201 (2011).

12. Carpene, E. *et al.* Dynamics of electron-magnon interaction and ultrafast demagnetization in thin iron films. *Phys. Rev. B - Condens. Matter Mater. Phys.* **78**, 1–6 (2008).

13. Eich, S. *et al.* Band structure evolution during the ultrafast ferromagnetic-paramagnetic phase transition in cobalt. *Sci. Adv.* **3**, 1–9 (2017).

14. Carpene, E., Hedayat, H., Boschini, F. & Dallera, C. Ultrafast demagnetization of metals: Collapsed exchange versus collective excitations. *Phys. Rev. B - Condens. Matter Mater. Phys.* **91**, 1–8 (2015).

15. Tengdin, P. *et al.* Critical behavior within 20 fs drives the out-of-equilibrium laser-induced magnetic phase transition in nickel. *Sci. Adv.* **4**, 1–9 (2018).

16. Kimling, J. *et al.* Ultrafast demagnetization of FePt:Cu thin films and the role of magnetic heat capacity. *Phys. Rev. B - Condens. Matter Mater. Phys.* **90**, 1–9 (2014).

17. Wilson, R. B. & Coh, S. Parametric dependence of hot electron relaxation time-scales on electron-electron and electron-phonon interaction strengths. *Commun. Phys.* **3**, (2020).

18. Gilmore, K., Idzerda, Y. U. & Stiles, M. D. Identification of the dominant precession-damping mechanism in Fe, Co, and Ni by first-principles calculations. *Phys. Rev. Lett.* **99**, 1–4 (2007).

19. Kamberský, V. On the Landau–Lifshitz relaxation in ferromagnetic metals. *Can. J. Phys.* **48**, 2906–2911 (1970).

20. Haag, M., Illg, C. & Fähnle, M. Role of electron-magnon scatterings in ultrafast demagnetization. *Phys. Rev. B - Condens. Matter Mater. Phys.* **90**, 1–6 (2014).

21. Tveten, E. G., Brataas, A. & Tserkovnyak, Y. Electron-magnon scattering in magnetic heterostructures far out of equilibrium. *Phys. Rev. B - Condens. Matter Mater. Phys.* **92**, 1–5 (2015).

22. Farle, M. Ferromagnetic resonance of ultrathin metallic layers. *Reports Prog. Phys.* **61**, 755–826 (1998).

23. Schoen, M. A. W. *et al.* Magnetic properties in ultrathin 3d transition-metal binary alloys. II. Experimental verification of quantitative theories of damping and spin pumping. *Phys. Rev. B* **95**, 1–9 (2017).

24. Kuneš, J. & Kamberský, V. First-principles investigation of the damping of fast magnetization precession in ferromagnetic (formula presented) metals. *Phys. Rev. B - Condens. Matter Mater. Phys.* **65**, 1–3 (2002).

25. Fähnle, M. & Steiauf, D. Breathing Fermi surface model for noncollinear magnetization: A generalization of the Gilbert equation. *Phys. Rev. B - Condens. Matter Mater. Phys.* **73**, 1–5 (2006).

26. Allen, P. B. Empirical electron-phonon values from resistivity of cubic metallic elements. *Phys. Rev. B* **36**, 2920–2923 (1987).

27. Koopmans, B., Ruigrok, J. J. M., Dalla Longa, F. & De Jonge, W. J. M. Unifying ultrafast magnetization dynamics. *Phys. Rev. Lett.* **95**, 1–4 (2005).

28. Zhang, W. *et al.* Unifying ultrafast demagnetization and intrinsic Gilbert damping in Co/Ni bilayers with electronic relaxation near the Fermi surface. *Phys. Rev. B* **96**, 1–7 (2017).

29. Mathias, S. *et al.* Probing the time-scale of the exchange interaction in a ferromagnetic alloy. *Proc. Natl. Acad. Sci. U. S. A.* **109**, 4792–4797 (2012).

30. Brorson, S. D. *et al.* Femtosecond room-temperature measurement of the electron-phonon coupling constant in metallic superconductors. *Phys. Rev. Lett.* **64**, 2172–2175 (1990).

31. Gloskovskii, A. *et al.* Electron emission from films of Ag and Au nanoparticles excited by a femtosecond pump-probe laser. *Phys. Rev. B - Condens. Matter Mater. Phys.* **77**, 1–11 (2008).

32. Chan, W. L., Averback, R. S., Cahill, D. G. & Lagoutchev, A. Dynamics of femtosecond laser-induced melting of silver. *Phys. Rev. B - Condens. Matter Mater. Phys.* **78**, 1–8 (2008).

33. Atxitia, U., Ostler, T. A., Chantrell, R. W. & Chubykalo-Fesenko, O. Optimal electron, phonon, and magnetic characteristics for low energy thermally induced magnetization switching. *Appl. Phys. Lett.* **107**, (2015).

34. Jakobs, F. *et al.* Unifying femtosecond and picosecond single-pulse magnetic switching in Gd-Fe-Co. *Phys. Rev. B* **103**, 18–22 (2021).

35. Davies, C. S. *et al.* Pathways for Single-Shot All-Optical Switching of Magnetization in Ferrimagnets. *Phys. Rev. Appl.* **13**, 1 (2020).

36. Ostler, T. A. *et al.* Ultrafast heating as a sufficient stimulus for magnetization reversal in a ferrimagnet. *Nat. Commun.* **3**, (2012).

37. Ceballos, A. *et al.* Role of element-specific damping in ultrafast, helicity-independent, all-optical switching dynamics in amorphous (Gd,Tb)Co thin films. *Phys. Rev. B* **103**, 24438 (2021).

38. Maier, W. F., Stowe, K. & Sieg, S. Combinatorial and high-throughput materials science. *Angew. Chemie - Int. Ed.* **46**, 6016–6067 (2007).

39. Geng, J., Nlebedim, I. C., Besser, M. F., Simsek, E. & Ott, R. T. Bulk Combinatorial Synthesis and High Throughput Characterization for Rapid Assessment of Magnetic Materials: Application of Laser Engineered Net Shaping (LENS[TM]). *JOM* **68**, 1972–1977 (2016).

40. Koinuma, H. & Takeuchi, I. Combinatorial solid-state chemistry of inorganic materials. *Nature Materials* **3**, 429–438 (2004).

41. Takeuchi, I., Lauterbach, J. & Fasolka, M. J. Combinatorial materials synthesis. *Mater. Today* **8**, 18–26 (2005).

42. Hohensee, G. T., Hsieh, W. P., Losego, M. D. & Cahill, D. G. Interpreting picosecond acoustics in the case of low interface stiffness. *Rev. Sci. Instrum.* **83**, (2012).

43. Cahill, D. G. Analysis of heat flow in layered structures for time-domain thermoreflectance. *Rev. Sci. Instrum.* **75**, 5119–5122 (2004).

44. Jiang, P., Qian, X. & Yang, R. Tutorial: Time-domain thermoreflectance (TDTR) for thermal property characterization of bulk and thin film materials. *J. Appl. Phys.* **124**, (2018).

45. Gomez, M. J., Liu, K., Lee, J. G. & Wilson, R. B. High sensitivity pump-probe measurements of magnetic, thermal, and acoustic phenomena with a spectrally tunable oscillator. *Rev. Sci. Instrum.* **91**, (2020).

46. Kang, K., Koh, Y. K., Chiritescu, C., Zheng, X. & Cahill, D. G. Two-tint pump-probe measurements using a femtosecond laser oscillator and sharp-edged optical filters. *Rev. Sci. Instrum.* **79**, (2008).

47. Johnson, P. B. & Christy, R. W. Optical constants of transition metals. *Phys. Rev. B* **9**, 5056–5070 (1974).

48. P. B. Johnson and R. W. Christy. Optical Constant of the Nobel Metals. *Phys. Rev. B* **6**,

4370–4379 (1972).

49. Ordal, M. A., Bell, R. J., Alexander, R. W., Newquist, L. A. & Querry, M. R. Optical properties of Al, Fe, Ti, Ta, W, and Mo at submillimeter wavelengths. *Appl. Opt.* **27**, 1203 (1988).

**Acknowledgements**

The work by R. M., V. H. O, and R. B. W. was primarily supported by the U.S. Army Research Laboratory and the U.S. Army Research Office under contract/grant number W911NF-18-1-0364 and W911NF-20-1-0274. R. M. and R. B. W. also acknowledge support by NSF (CBET – 1847632). The work by L. V. and S. C. was supported by the U.S. Army Research Laboratory and U.S. Army Research Office under contract/grant number W911NF-20-1-0274. Energy Dispersive X-Ray Spectroscopy (EDS) analyses were performed at the Central Facility for Advanced Microscopy and Microanalysis (CFAMM) at UC Riverside.

**Author Contributions**

R. M. and R. B. W. designed the experiments. R. M. prepared all the samples and characterized them, and performed TR-MOKE experiments. V. H. O performed VSM measurements. L. V. performed hierarchical clustering analyses. S. C. performed DFT calculations. R. M. and R. B. W. analyzed the data and wrote the manuscript, with discussions and contributions from L. V. and S. C.

**Additional Information:** Supplementary information is provided with this manuscript.

**Competing Interests:** The authors declare no competing interest.

**Data Availability:** The data that supports the findings of this paper are available from the corresponding author upon reasonable request.

**Correspondence:** Correspondence and request for additional information must be addressed to rwilson@ucr.edu